# Modeling the magnetoresistance vs. field curves of GMR multilayers with antiferromagnetic and/or orthogonal coupling by assuming single-domain state and coherent rotation

K. Szász and I. Bakonyi[*]

*Wigner Research Centre for Physics, Hungarian Academy of Sciences.*
*H-1121 Budapest, Konkoly-Thege útr 29-33, Hungary*

**Abstract** ─ In order to better understand the role of possible couplings in determining the giant magnetoresistance (GMR) behavior of multilayers, a knowledge of the dependence of the *GMR* on magnetic field *H* appears to be useful. Since a few specific cases have only been treated theoretically in the literature, it was decided to carry out a modeling of the *GMR*(*H*) curves of ferromagnetic/non-magnetic (FM/NM) multilayers with various interlayer couplings. For simplicity, we focused on a trilayer structure ($FM_1/NM/FM_2$) corresponding fairly well to the case of a large number of FM/NM bilayers. To carry out the calculations, some fundamental assumptions were made: (i) each FM layer consists of a single domain and the magnetizations are in the layer planes; (ii) the magnetization of each layer is the same; (iii) the magnetization vectors rotate in the plane of the layers in an external magnetic field. In order to calculate the *GMR*(*H*) function, we need to know the magnetization process in the trilayer, i.e., the *M*(*H*) function. Therefore, first we calculate the equilibrium angle $\varphi(H)$ between the two magnetization vectors as a function of the field by minimizing the total energy of the multilayer. According to most previous theoretical and experimental works, the angular dependence of the *GMR* is fairly well described by the relation $GMR(\varphi) \propto (1 - \cos \varphi)$ and we used this relation to derive the *GMR*(*H*) function. Along this line, the *M*(*H*) and *GMR*(*H*) curves were calculated for the following cases: (i) pure AF coupling; (ii) pure orthogonal coupling; (iii) AF coupling and orthogonal coupling simultaneously present. As to the calculation of the *GMR*(*H*) curves, some of these configurations have not yet been treated formerly or for some specific parameter values only. For those cases for which calculations were reported in the literature for *M*(*H*) and *GMR*(*H*), our results agree with previous reports.

---

[*] Corresponding author. E-mail: bakonyi.imre@wigner.mta.hu

# 1. INTRODUCTION

In recent years, new devices produced by nanofabrication routes have been commercialized which operate by using the two possible spin states (*spin-up* and *spin-down*) of electrons. The appearance of such nanoscale devices has laid the foundation of *spin-electronics* (or *spintronics*),[1-3] an industry of broad future perspective. This progress has been made possible via the rapid development of thin film technologies for the preparation of nanoscale metallic layered structures (*multilayers*) which can exhibit electrical transport properties significantly different from those of bulk materials. This can occur if the thickness of constituent layers becomes smaller than the characteristic length scales of electron transport (e.g., mean free path, spin diffusion length). In case one of the layers is ferromagnetic (FM) and the magnetization orientation changes on a scale smaller than these characteristic lengths, so-called *spin-dependent electron transport* phenomena different from those known in bulk materials may also arise due to the polarisability of conduction electrons when they travel between two neighboring FM layers through a non-magnetic (NM) spacer layer. An outstanding example of such phenomena is the *giant magnetoresistance (GMR)* effect discovered in FM/NM multilayer films[4,5] in the late 1980's. Magnetoresistive sensors utilizing the GMR effect have already found widespread applications, e.g., for measuring or detecting a magnetic field.[6] In present-day magnetic hard disk drives, the read-out heads operate mostly with the GMR principle. Their introduction made it possible to maintain a very high rate of magnetic storage density increase[7] and GMR performance ensures that this trend can continue even in the future.

It was demonstrated in the original reports on the GMR[4,5] that the large resistance change upon the application of an external magnetic field arose because in zero magnetic field the equilibrium state of the FM/NM multilayers investigated was an antiparallel alignment of the neighboring layer magnetizations. Such a large nanoscale variation of the magnetization orientation in zero external field when going from one magnetic layer to the adjacent one results in much stronger electron scattering, i.e., larger resistivity, in comparison with the magnetically homogeneous, saturated state at high magnetic fields which state exhibits a significantly lower electrical resistivity. The antiparallel alignment was the consequence of a strong antiferromagnetic (AF) coupling[8,9] mediated by the non-magnetic spacer layer between the FM layer magnetizations.

It was subsequently soon discovered that in granular magnetic alloys[10,11] which contain nanoscale superparamagnetic (SPM) particles embedded in a NM matrix, electron pathways between the SPM entities with fluctuating magnetic moment orientations result in a similar spin-dependent scattering as that yielding the GMR effect in FM/NM multilayers. It became clear from this observation that a necessary condition for the occurrence of the phenomenon of GMR is not the AF coupling but some degree of antiparallel alignment of the neighboring layer magnetizations. Of course, a strong AF coupling ensures the largest degree of alignment, i.e., a completely antiparallel state. The presence of another type of coupling or various magnetic anisotropies may also play a role in determining what the equilibrium configuration of the sublayer magnetizations is at $H = 0$. A phenomenological discussion of the possible interlayer couplings was treated by Schäfer et al.[12] whereas the magnetic anisotropies playing a decisive role in multilayers were summarized by Johnson et al.[13]

The GMR magnitude is governed by the misalignment of the neighboring layer magnetizations in zero field since the larger the misalignment, the larger the excess resistivity with respect to the saturated, high-field state. Namely, the degree of spin-dependent scattering when an electron travels through a non-magnetic spacer between two non-aligned neighboring magnetic regions depends on the angle $\varphi$ between the two magnetization vectors.

In this manner, a knowledge of the zero-field magnetic configuration of the sublayer magnetizations for various interlayer couplings and magnetic anisotropies in FM/NM multilayers is of crucial importance. For a given type of coupling and/or anisotropy, the equilibrium ($H = 0$) state can be theoretically calculated (at least for the case of a homogeneous magnetization which is equivalent to a single-domain state in each layer). Furthermore, by assuming a given type of magnetization process, also the field evolution of the relative orientations of the neighboring layer magnetizations, i.e., the $\varphi(H)$ function can be determined from which, then, the magnetization curve $M(H)$ can be derived. In the absence of any coupling, the sublayer magnetization orientations are determined solely by the magnetic anisotropies present, including demagnetization effects. Then, a random orientation of the neighboring layer magnetizations can also lead to a sufficient misalignment and to a significant GMR effect. In case we know the $GMR(\varphi)$ function, we can also determine the field-evolution of the magnetoresistance, i.e., the $GMR(H)$ curve.

It should be noted at this point that our original motivation for dealing with the shape of the $GMR(H)$ curves came from former studies on the GMR of electrodeposited multilayers. Namely, it has been a common observation[14] that the $MR(H)$ curves of electrodeposited multilayers often exhibit a strongly non-saturating behavior, in some cases without a sign of approach to saturation in magnetic fields as high as 10 kOe. It could be shown[15] that the strongly non-saturating behavior can be ascribed to the presence of SPM regions in the magnetic layers. By using a decomposition procedure,[15] it is possible to separate the SPM contribution ($GMR_{SPM}$) from the conventional ferromagnetic GMR contribution ($GMR_{FM}$) which latter is the sole $GMR$ term in FM/NM multilayers with magnetic layers exhibiting purely ferromagnetic characteristics.

Furthermore, it has been shown in a recent work[16] that for electrodeposited Co/Cu multilayers, the $GMR_{FM}$ term when properly extracted from the experimental $GMR$ data does not show an oscillatory behavior as a function of the spacer layer (Cu) thickness although this feature is well documented for multilayers prepared by physical deposition methods (for the Co/Cu multilayer system, see Refs. 17-19). In electrodeposited Co/Cu multilayers for Cu layer thicknesses above 1 to 2 nm, the $GMR_{FM}$ term was found[16] to increase smoothly with Cu layer thickness up to a maximum at about 4 nm. This fact, together with the large relative remanence (~0.75) of the magnetization in these multilayers,[16] suggests the absence of any significant antiferromagnetic (AF) coupling in electrodeposited Co/Cu multilayers. As noted already above, in an uncoupled state when the magnetization orientation distribution is more or less random, there is some degree of antiparallel alignment of the neighboring layer magnetizations which is a sufficient condition for the occurrence of a GMR effect.

It appeared, therefore, that in order to better understand the evolution of GMR with spacer thickness and the role of possible couplings in multilayers, especially the behavior observed in electrodeposited multilayers, a knowledge of the dependence of $GMR$ on magnetic field $H$ would be useful.

As a general guideline to the information content of the shape of the $GMR(H)$ curves, Fert and Bruno[20] provided a brief summary. According to their discussion, the ideal case of a perfect antiparallel alignment in zero external field (which is realized if there is a strong AF coupling between the magnetic layers whereas magnetic anisotropy and coercivity is negligible) results in a typical bell-shaped curve for which the decrease of resistivity from the value at $H = 0$ starts as $(H/H_s)^2$ where $H_s$ is the saturation field against the AF coupling. The experimentally measured $MR(H)$ curve for a Co/Cr multilayer is provided by Fert and Bruno[20] as evidencing this ideal situation. They also make the qualitative notion that if the antiparallel alignment is not perfect at $H = 0$, a sharp maximum of the $MR(H)$ curve can be observed, as demonstrated by the data shown for a sputtered Fe/Cu multilayer.[20]

There have already been several reports in the literature[21-33] on the theoretical calculation of the $GMR(H)$ curves of coupled multilayers. The first work by Folkerts[21] deals with an antiferromagnetically coupled system with cubic symmetry or uniaxial anisotropy. The cases of both the cubic and the uniaxial symmetry have subcases depending on the direction of the external magnetic field with respect to the crystalline directions. Subsequently, Fujiwara and Parker[22] presented an analytical model for finite number of layers in case of bilinear and biquadratic exchange with uniaxial anisotropy which model was solved for some specific values of the coupling constants and the anisotropy constant only. Several further reports have been published[23,24,26,27,32] which rely on this model.

A very comprehensive study of the possible magnetic configurations of exchange-coupled layers was carried out by Ustinov et al.[25] who then also calculated the $GMR(H)$ curves for some specific cases. Finally, we mention the works by Reiss and coworkers[30,31,33] who modeled the $GMR(H)$ curves for Permalloy/Cu multilayers, with specific attention to stacks with an alternating sequence of thick and thin Cu layers which provided an alternation of strong and weak coupling between adjacent Permalloy layers.

In spite of the significant previous efforts, we can establish that not all possible combinations of interlayer couplings and anisotropy have been treated in sufficient detail or at most for some specific configurations and parameter values only. It could also not always be properly inferred from the reports what particular $GMR(\varphi)$ relation was used (or, if specified, the choice was not properly justified) for the determination of the $GMR(H)$ curve.

Therefore, in order to extract some additional information on the interlayer couplings (or on their absence) from the experimental magnetoresistance data, we have carried out a modeling of the $GMR(H)$ curves of FM/NM multilayers with various interlayer couplings and anisotropies. To achieve this goal, we need to know the $M(H)$ curve first for a given combination of interlayer couplings and anisotropies. Of course, there is a vast amount of literature on calculating the equilibrium magnetic states of multilayers and their magnetization curves[21,24,25,27,28,32,34-43] and the aim was not to reproduce these results. However, in these previous reports the $M(H)$ curves were only presented and not the $\varphi(H)$ function which is required for deriving the $GMR(H)$ curves in a manner as will be detailed in Section II. Therefore, we needed to recalculate the equilibrium magnetic configurations as a function of the external magnetic field [i.e., to establish the $\varphi(H)$ function] for each specific coupling and anisotropy of interest here.

Corresponding to the above discussion, the purpose of our study was a modeling of the $GMR(H)$ curves of FM/NM multilayers for various couplings and magnetic anisotropies. Considerations will be restricted to cases corresponding to the assumption of a single-domain state and a coherent rotation of the spins in each magnetic layer during the magnetization reversal process. In those cases where we carried out the calculation for a configuration treated already beforehand in the literature, a good agreement with previous results was obtained.

In the present work, the results of modeling the $GMR(H)$ curves will be described for the case of
(i)    pure AF coupling,
(ii)   pure orthogonal coupling and
(iii)   simultaneous presence of AF coupling and orthogonal coupling.

The combinations of various couplings and the presence of a uniaxial anisotropy will be treated in a subsequent paper.

It should be noted that, as expected, no hysteresis in the $M(H)$ and $GMR(H)$ curves occurs in the absence of a magnetic anisotropy.

Before presenting the results for the various couplings to be investigated in the current paper, Section 2 first describes the procedures and conditions applied for calculating the equilibrium angle between adjacent layer magnetizations and the field dependence of the *GMR* magnitude.

## 2. METHODS OF CALCULATION
### 2.1. Calculation of the field dependence *φ*(*H*) of the equilibrium angle

For simplicity, we focus on a trilayer structure ($FM_1/NM/FM_2$), following the treatment applied for studying the *M*(*H*) curves for various couplings and anisotropies by Dieny et al.[35] These authors have pointed out that although this simplified treatment represents a restricted case, it still contains much of the physics of the magnetization processes of FM/NM multilayers. For a number of cases, they have also extended the model consideration to study the influence of the number of layers on the magnetization processes.[35]

To carry out the calculations, some fundamental assumptions are made: (i) each FM layer consists of a single domain and the magnetizations are in the layer planes; (ii) the magnetization and thickness of each magnetic layer is the same ($|\mathbf{M}_1| = |\mathbf{M}_2| = M$, $d_1 = d_2 = d$) and (iii) the magnetization vectors rotate in the plane of the layers in an external magnetic field.

In order to calculate the *GMR*(*H*) curve, we need to know the magnetization process in the trilayer structure, i.e., the *M*(*H*) function. Therefore, first we calculate the equilibrium angle *φ*(*H*) between the two magnetization vectors for a given value of the magnetic field *H* by minimizing the total magnetic energy of the multilayer. A knowledge of the equilibrium angle *φ* between $\mathbf{M}_1$ and $\mathbf{M}_2$ as a function of *H* enables us to obtain the *M*(*H*) curve. Afterwards, if the *GMR*(*φ*) relation is known, the *GMR*(*H*) function can already be derived.

In the absence of a magnetic anisotropy, the magnetic areal energy-density of the trilayer structure $FM_1/NM/FM_2$ can be written as the sum of two contributions:

$$E = E_{magn} + E_{coupl}.$$

In the following, we shall express (i) the term $E_{magn}$ with the help of the angles $\vartheta_1$ and $\vartheta_2$ between the external magnetic field and the magnetization vectors as sketched in Fig. 1 and (ii) the term $E_{coupl}$ with the help of the angle $\varphi = \vartheta_1 + \vartheta_2$.

For each magnetic layer (i = 1 and 2), the magnetic (Zeeman) energy-density term is of the form $E_{magn} = -HMd\cos\vartheta_i$ where $\vartheta_i$ is the angle between the magnetization and the magnetic field vector and *d* is the magnetic layer thickness.

For the coupling between the magnetizations of the two magnetic layers in the trilayer structure, we may start from the phenomenological description of the interlayer coupling as formulated by Schäfer et al.[12] who expanded the coupling energy according to the powers of the product $\mathbf{M}_1 \cdot \mathbf{M}_2$.

The first-order term in the expansion of coupling yields the coupling energy-density $E_{coupl} = -J_1 \cos\varphi$ per unit area where $\varphi = \vartheta_1 + \vartheta_2$ is the angle between $\mathbf{M}_1$ and $\mathbf{M}_2$ and $J_1$ is the coupling constant (per unit area) of this so-called bilinear coupling. The antiparallel alignment of $\mathbf{M}_1$ and $\mathbf{M}_2$ corresponds to $\varphi = 180°$. This state will be an energy minimum if $J_1 < 0$ and this describes an AF coupling. It should be noted that for $J_1 > 0$, a FM coupling between the layer magnetizations is obtained.

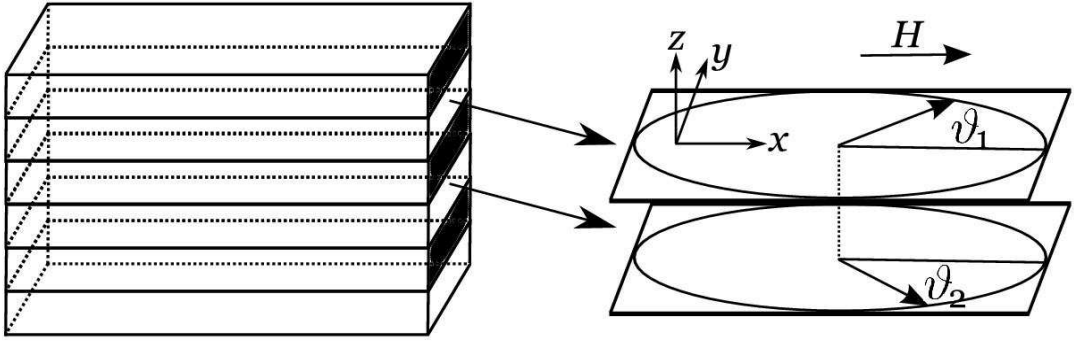

***Fig. 1*** *The sketch of a multilayer with details of the configuration concerning two adjacent magnetic layers. In the x-y plane, the lateral size of the layers is macroscopic while in the z direction the layers are a few nanometers thick only. The external field H is directed along the x axis. The thickness of the magnetic layers is taken as d whereas $\vartheta_1$ and $\vartheta_2$ denote the angle between the magnetic field direction and the layer magnetization in two adjacent magnetic layers.*

The second-order term in the expansion is proportional to $(M_1 \cdot M_2)^2$ and, therefore, it is often called as biquadratic coupling. The resulting coupling energy-density per unit area is $E_{coupl} = -\frac{1}{2} J_2 \cos^2 \varphi$ where $J_2$ is the coupling constant referred to a unit area (the prefactor ½ was introduced for later convenience). This energy can be a minimum (stable state) with $J_2 < 0$ if $\varphi = 90º$ and this yields an orthogonal coupling or 90º-coupling. Depending on the relative magnitude of $J_1$ and $J_2$, the resulting coupling may be dominated by either the AF or the orthogonal coupling or, if the strengths of the two couplings are comparable, a non-collinear alignment with $90º < \varphi < 180º$ may realize.

Having set the formalism for the current situation of a trilayer, the task is now to find the equilibrium value of the angles $\vartheta_1$ and $\vartheta_2$ for a given magnetic field strength which angles correspond to the minimum of the total energy. To get the energy minimum, the first derivative (vector) must be zero and the second derivative (or the determinant and the 1,1 element of the second derivative matrix) must be greater than zero:

$$\frac{\partial E}{\partial \vartheta_i} = 0 \text{ and } \frac{\partial^2 E}{\partial \vartheta_i^2} > 0 \qquad (i = 1, 2).$$

(There is another formulation of the statement concerning the local minimum: the eigenvalues of the second derivative matrix have to be greater than zero. It is easy to show that the two statements are equivalent.)

The solution of these equations describes the magnetization process, and, thus the $\varphi(H)$ function yielding the equilibrium angle $\varphi$ between the two magnetizations as a function of the external magnetic field $H$.

### 2.2. Calculation of the GMR(H) curves

As to the dependence of the *GMR* on the angle $\varphi (= \vartheta_1 + \vartheta_2)$ between the orientations of the adjacent layer magnetization vectors, there have been several theoretical considerations on this issue[44-51] by using various approaches and treatments as well as specific material parameters including layer thicknesses. We will only consider works where the current-in-plane (CIP) geometry was considered. The basic result from these theoretical studies is that the angular dependence of the *GMR* magnitude is in most cases described by the relation *GMR($\varphi$)* $\propto$ (1 - cos $\varphi$) although serious deviations can be expected from this behavior if certain conditions are not fulfilled. In most of these theoretical studies,[44-48,50] not a specific FM/NM material combination was investigated but the treatment was formulated for a multilayer or trilayer structure fulfilling some requirements concerning the material properties. As an example from this group, we may mention the work of Barnas et al.[47] who considered a $FM_1$/NM/$FM_2$ trilayer (by imposing perfectly reflecting boundary conditions, the results obtained are valid for a corresponding superlattice as well). Spin-dependent electron scattering on impurities as well as on interfacial roughness was taken into account. In the quasiclassical limit and in the free-electron approximation for the conduction electrons, the angular variation of the GMR magnitude was very close to the relation *GMR($\varphi$)* $\propto$ (1 - cos $\varphi$) both for a uniform crystal potential and for the case of large spin-dependent potentials. In the quantum limit, i.e., in the presence of strong quantum interference of electron waves reflected from the interfaces and/or outer surfaces, the cos $\varphi$ behavior occurs only if the trilayer structure is symmetrical (both the thicknesses of the magnetic layers and the scattering potentials of the two interfaces are the same) and, simultanouesly, the crystal potential should be independent of the spin orientation. In the case of either an asymmetrical trilayer structure or a spin-dependent electron potential, the quantum limit yields significant deviation from the relation *GMR($\varphi$)* $\propto$ (1 - cos $\varphi$).

In the other group of theoretical studies on the angular dependence of the *GMR*,[49,51] in both cases a Co/CuCo trilayer structure was considered and, first, a self-consistent calculation of the electronic band structure was carried by using the Korringa-Kohn-Rostoker (KKR) method and, finally, the Kubo-Greenwood formula was used for calculating the conductivity. Brown et al.[49] neglected the spin-orbit coupling and the electron scattering by impurities, phonons and magnons was modeled with a layer- and spin-dependent complex self energy. In this particular case, the result was that the deviation from the relation *GMR($\varphi$)* $\propto$ (1 - cos $\varphi$) was significant (it amounted to a reduction by 30 % at $\varphi$ = 90º). By contrast, Blaas et al.[51] applied a fully-relativistic spin-polarized screened KKR method for the electronic band structure calculation to account for the spin-orbit coupling and for the spin-dependent scattering and used the coherent potential approximation for describing the scattering. They also considered the intermixing at the interfaces and calculated simultaneously the angular dependence of the interlayer exchange coupling as well. The magnitude of the *GMR* was calculated directly for nine values of $\varphi$ in the range 0º < $\varphi$ < 180º with the *GMR* being zero and maximum at $\varphi$ = 0º and $\varphi$ = 180º, respectively. An angular dependence of the *GMR* very similar to the relation *GMR($\varphi$)* $\propto$ (1 - cos $\varphi$) was obtained: the calculated values could be very well fitted by the function *GMR($\varphi$)* = $a_1$(1 - cos $\varphi$) + $a_2$(1 – $\cos^2 \varphi$) whereby $a_1$ was larger than $a_2$ almost by an order of magnitude.

From the experimental side, the angular variation of the *GMR* was found to be very close to the relation *GMR($\varphi$)* $\propto$ (1 - cos $\varphi$) in the CIP geometry for various systems: Fe/Cr/Fe (Ref. 52) and Co/Cu/NiFe (Ref. 53) trilayer sandwiches, NiFe/Cu/NiFe/FeMn spin valve[54] as well as NiFe/Ag (Ref. 55) and NiFe/Ag/Co/Ag and NiFe/Cu/Co/Cu (Ref. 56) multilayers.

From these works, we can conclude that the *GMR($\varphi$)* function is fairly well approximated by the relation *GMR($\varphi$)* $\propto$ (1 - cos $\varphi$) and this functional form will be used in the following. In some cases, we shall calculate the *GMR(H)* curves for comparison also by taking into account the deviation from the relation *GMR($\varphi$)* $\propto$ (1 - cos $\varphi$) as calculated by Blaas et al.[51]

This way, for a given magnetic field *H* if we know the equilibrium angle *$\varphi(H)$* between the two magnetic vectors, from this we can get the *GMR* value for any *φ(H)*, thus we can determine the *GMR(H)* function.

Along this line, the *M(H)* and *GMR(H)* curves were calculated for the following cases:
(i) pure AF coupling;
(ii) pure orthogonal coupling;
(iii) AF coupling and orthogonal coupling simultaneously present.

The results of modeling by considering also the presence of a uniaxial anisotropy will be presented in a subsequent paper.

As noted in Section 2.1, the AF (or bilinear) coupling and the orthogonal (or biquadratic) coupling can be phenomenologically derived as the first two terms in the expansion according to the powers of the product $\boldsymbol{M}_1 \cdot \boldsymbol{M}_2$. The presence of an AF exchange coupling between two ferromagnetic layers via a non-magnetic spacer layer has been demonstrated by Grünberg et al.[57] It could be shown subsequently[58] that, under proper conditions, an orthogonal coupling may also occur between adjacent layer magnetizations. Later on, various microscopic models[59-62] have been put forward to explain the physical origin of the orthogonal coupling. In addition to these models providing a direct mechanism for this type of coupling, Marrows and Hickey[29] suggested an indirect way how the orthogonal coupling may be realized by assuming a lateral alternation of areas with either AF or FM coupling, the latter created, e.g., by localized defects such as pinholes[63] or by layer thickness fluctuations[64] in the spacer material. Namely, when the fluctuation length scale of the coupling is comparable to or smaller than the domain wall width of the magnetic layer material, the frustration between the AF and FM couplings may lead to an orthogonal coupling since in this case both magnetic layers can maintain a homogeneously magnetized low-energy state and simultaneously avoiding the coupling frustration. It is interesting to note finally that the atomic step microscopic model of the orthogonal coupling by Demokritov et al.[60] was extended[65,66] to show that atomic step irregularities at the FM/NM layer interfaces can yield either AF or FM coupling, depending on the relative positions of the step at the interfaces opposite across the spacer. This way we can close the loop since this provides another microscopic mechanism, in terms of the fluctuation model of Marrows and Hickey[29] for the occurrence of an orthogonal coupling.

All these considerations justify the importance of the orthogonal coupling which should, therefore, be treated on equal footing with the AF coupling when evaluating the possible magnetic configurations of magnetic layers.

## 3. RESULTS AND DISCUSSION
### 3.1. The case of pure AF coupling

In the absence of a magnetic anisotropy, the magnetization vectors of the two magnetic layers have the same $\vartheta$ angle symmetrically with respect to the direction of the magnetic field ($\vartheta_1 = \vartheta_2 = \vartheta = \varphi/2$). The areal energy-density of the trilayer is then given by

$$E(\vartheta) = -2MHd\cos\vartheta - J_1\cos(2\vartheta).$$

From the first derivative, we get

$$MHd\sin\vartheta + 2J_1\sin\vartheta\cos\vartheta = 0.$$

The solutions of this equation provide first the saturation state. There is a minimum in the energy if

$$H > H_s = -\frac{2J_1}{Md}$$

where $H_s$ is the saturation field and $J_1 < 0$ for the case of AF coupling. The second solution describes the magnetization process:

$$\cos\vartheta = -\frac{MHd}{2J_1} = \frac{H}{H_s}.$$

Since in this expression the second derivative is always greater than zero, it provides the equilibrium angles between one of the magnetizations and the field for a given magnetic field. By making use of the relation $\vartheta = \varphi/2$, the equilibrium angle between the two magnetizations varies with magnetic field as

$$\varphi = 2\arccos\frac{H}{H_s}.$$

The function $\varphi(H)$ is displayed in Fig. 2.

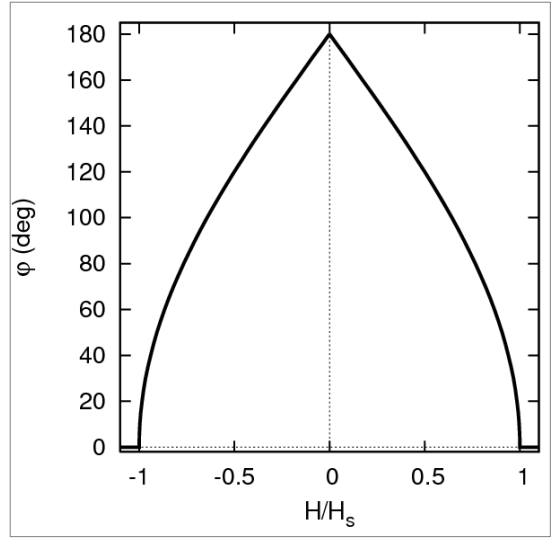

***Fig. 2*** *Dependence of the equilibrium angle $\varphi$ between the two magnetization vectors on the magnetic field H normalized with the saturation field $H_s$ for a trilayer with pure AF coupling.*

On the basis of the above equations, the net magnetization is given by

$$\frac{M}{M_s} = \cos\vartheta,$$

where $M_s = 2M$ is the saturation magnetization. Thus, the resulting magnetization curve $M(H)$ is linear (Fig. 3a). On the other hand, the $GMR(H)$ curve has the characteristic bell-shaped form (see Fig. 3b) as shown by the experimental results of Fert and Bruno[20] (see Section 1) for the case of AF coupling with negligible anisotropy. It can be observed in Fig. 3b that the $GMR(H)$ curve changes only slightly if, in addition to the $(1 - \cos\varphi)$ term, also the quadratic term $(1 - \cos^2\varphi)$ as calculated by Blaas et al.[51] is taken into account for describing the angular dependence of the GMR.

Evidently, AF coupling without magnetic anisotropy does not result in a hysteresis of either the $M(H)$ or the $GMR(H)$ curves.

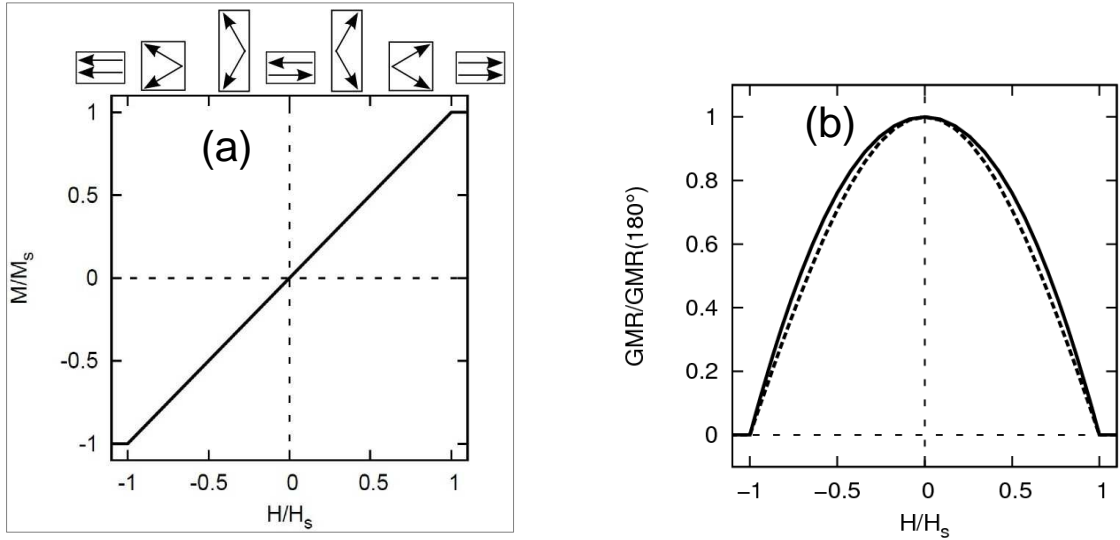

*Fig. 3* The magnetization curve (a) and the GMR(H) curve (b) for a trilayer with pure AF coupling. The M(H) curve was normalized with the saturation magnetization $M_s$, the GMR(H) curve with the GMR value of the antiparallel state, GMR(180º), whereas the magnetic field was normalized with the saturation field $H_s$. In (a), the arrows in the small boxes show the alignment of the magnetization vectors. In (b), the dashed line shows the field dependence of the GMR by assuming the relation GMR($\varphi$) $\propto$ (1 - cos $\varphi$) for the angular dependence of the GMR whereas the solid line is for the case when the quadratic correction[51] to GMR($\varphi$) is also taken into account.

### 3.2. The case of pure orthogonal coupling

In the lack of a magnetic anisotropy, the magnetization vectors lie symmetrically with respect to $H$ ($\vartheta_1 = \vartheta_2 = \vartheta = \varphi/2$). The areal energy density of a FM/NM/FM trilayer is

$$E(\vartheta) = -2MHd \cos\vartheta - \frac{1}{2} J_2 \cos^2(2\vartheta).$$

The first derivative after reducing:

$$\left(MHd - 2J_2 \cos\vartheta + 4J_2 \cos^3\vartheta\right)\sin\vartheta = 0. \quad (1)$$

The first solution of this equation is $\sin\vartheta = 0$, which means the saturation state because if we write back this result into the second derivative, we get that this solution is energetically favorable when $H > H_s$ where $H_s = -2J_2/(Md)$ is the saturation field with $J_2 < 0$. In the absence of a magnetic field, the magnetization vectors are orthogonal to each other ($\vartheta = 45°$, $\varphi = 90°$) since an orthogonal coupling is present [this can also be inferred from eq. (1)].

From the other solution of eq. (1), we get the expression

$$\frac{H}{H_s} = 2\cos^3\vartheta - \cos\vartheta. \quad (2)$$

In this case, the second derivative will be greater than zero if $0° < \vartheta < 66°$. That is, the energy has a minimum state in this angular interval. However, we should keep in mind that according to the physical picture about the magnetization process, we should look only for solutions in the magnetic field range $0 < H < H_s$. This is because when reducing the magnetic field from $+H_s$ to zero, in the remanence state ($H = 0$) there will be an orthogonal alignment of the two magnetizations ($\vartheta = 45°$, $\varphi = 90°$) with a finite remanence $+M_r/M_s = \sqrt{2}/2 = 0.7071$. By applying an infinitesimally small magnetic field along the negative magnetic field direction, the orthogonally coupled arrangement of the two magnetizations, due to the lack of a magnetic anisotropy present, immediately jumps over to align the vectorial resultant of the two magnetizations along the magnetic field direction (i.e., $+M_r$ changes discontinuously to $-M_r$). Afterwards, by increasing the magnetic field from the initial small negative value to $H_s$, the evolution of the magnetization will be described by the same function as was the case for the magnetic field range $0 < H < H_s$.

By using the relation $\vartheta = \varphi/2$, one can obtain the magnetic field dependence of the equilibrium angle $\varphi$ between the two magnetizations in the range $0 < H < H_s$ which is shown in Fig. 4.

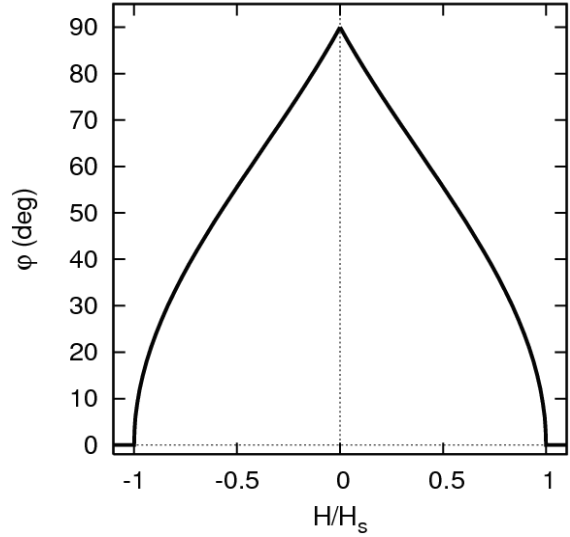

***Fig. 4*** *Dependence of the equilibrium angle $\varphi$ between the two magnetization vectors on the magnetic field H normalized with the saturation field $H_s$ for a trilayer with pure orthogonal coupling.*

The shape of the magnetization curve, i.e., the *M(H)* function, as derived from eq. (2), is displayed in Fig. 5a. From a knowledge of the dependence of $\varphi$ on magnetic field, the *GMR(H)* curve can be obtained as shown in Fig. 5b. It can be observed in Fig. 5 for the case of pure orthogonal coupling that, due to the different angular dependences of the magnetization and the *GMR*, the remanent magnetization decreases by about 30 % only with respect to the saturation magnetization, whereas the maximum *GMR* reduces by about 50 % with respect to the maximum possible antiparallel alignment ($\varphi = 180°$, pure AF coupling).

Similarly to Fig. 3b for the case of pure AF coupling, the dashed and solid lines in Fig. 5b show the *GMR(H)* curves without and with the quadratic correction of Blaas et al.[51], respectively, to the relation *GMR($\varphi$)* $\propto$ (1 - cos $\varphi$) describing the angular dependence of the *GMR*. It can bee seen that for the case of pure orthogonal coupling the shape of the *GMR(H)* curve remains unaltered by the quadratic correction just the magnitude of the maximum *GMR* is different for the two cases.

As in the previous case of AF coupling, an orthogonal coupling without magnetic anisotropy does not result in a hysteresis of either the *M(H)* or the *GMR(H)* curves.

Figure 5b also reveals that the *GMR(H)* curve exhibits a sharp peak at zero field and a downward curvature in the field dependence. This is in agreement with the qualitative remark of Fert and Bruno[20] for the case of a non-perfect antiparallel alignment at *H* = 0 as mentioned in the Section 1.

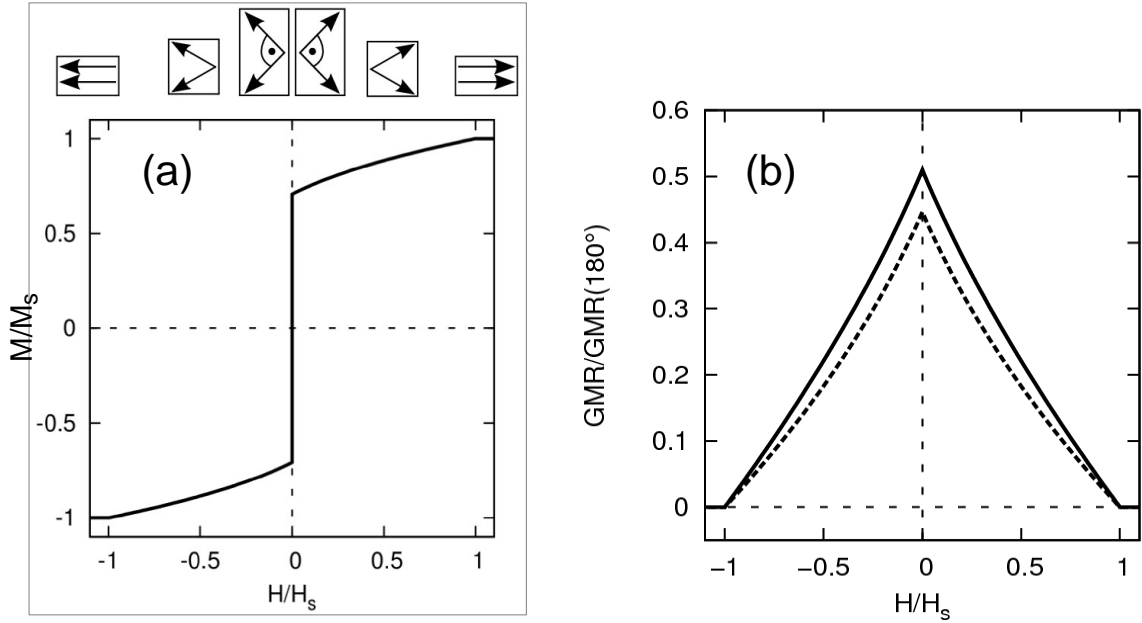

*Fig. 5* *(a) The normalized magnetization curve of an FM/NM/FM trilayer with pure orthogonal coupling. There is an irreversible magnetization jump at H = 0. The arrows in the small boxes show the alignment of the magnetization vectors; (b) The GMR(H) curve in the case of pure orthogonal coupling. Starting from saturation (parallel alignment), the resistivity increases until H = 0 where the angle between the two magnetizations is maximum (90 degree); upon reversing the magnetic field, the angle between the magnetizations decreases which results in a decrease of the resistivity. The dashed and solid lines show the results calculated similarly as explained at Fig. 3b.*

It should be noted that orthogonal coupling was assumed by Marrows and Hickey[29] to interpret their experimental results for the *M(H)* and *GMR(H)* curves on sputtered Co/Cu multilayers which were prepared under conditions to intentionally deteriorate locally the AF coupling between Co layers at some areas (series B in Ref. 29). This conclusion was deduced by model calculations in which they displayed the zero-field *MR* value against the remanence. Our above presented modeling results for the *M(H)* and *GMR(H)* curves of pure orthogonal coupling describe properly their experimental results. Not only the shape of the field-dependence of the magnetization and magnetoresistance curves matches very well for the calculated and experimental data but there is even good quantitative agreement of modeled and calculated remanence and zero-field magnetoresistance values (e.g., in both cases, the zero-field magnetoresistance drops to about half of the value of the fully AF-coupled state). This is a strong confirmation of the existence of a dominant orthogonal coupling in sample series B studied by Marrows and Hickey[29].

### 3.3. AF and orthogonal coupling simultaneously present

Due to the lack of a magnetic anisotropy, we can assume a symmetric arrangement with respect to *H* again ($\vartheta_1 = \vartheta_2 = \vartheta = \varphi/2$). The areal energy density of the trilayer is

$$E(\vartheta) = -2MHd \cos\vartheta - J_1 \cos(2\vartheta) - \frac{1}{2} J_2 \cos^2(2\vartheta).$$

From the first derivative, the following formulae can be derived.

The saturation field (both magnetization vectors are aligned along the external magnetic field) is given by the expression

$$H_s = -\frac{2J_1 + 2J_2}{Md}.$$

By introducing parameter $j = J_1/J_2 \geq 0$, the equation which describes the magnetization process is as follows:

$$\frac{H}{H_s} = \frac{j-1}{j+1}\cos\vartheta + \frac{2}{j+1}\cos^3\vartheta. \qquad (3)$$

First, it will be examined how the arrangement of the magnetization vectors depends on the value of $j$ in the absence of an external magnetic field ($H = 0$). If $j \geq 1$, it follows from eq. (3) that the magnetizations are aligned antiparallel. On the other hand, if $j < 1$, the equilibrium angle $\vartheta$ is determined by the equation:

$$\cos\vartheta = \sqrt{\frac{1-j}{2}}.$$

Accordingly, since $\vartheta = \varphi/2$ this means that whereas j varies from 0 towards 1, the angle $\varphi$ between the two magnetization vectors changes from 90º to 180º. The dependence of $\varphi$ on $j$ is illustrated in Fig. 6. It is a surprising fact that whereas one could naively expect a transition between the two couplings in a sense that $\varphi$ changes permanently from 90º to 180º as $J_1$ starts to dominate over $J_2$ (i.e., $j$ becomes larger and larger), the calculations reveal that the fully antiparallel state is reached already at a finite value of $j = 1$.

As a next step, it will be examined, from an inspection of the second derivative, how eq. (3) leads to an energy minimum for the above established two ranges of parameter $j$ and what will be the field dependence of the $M(H)$ and $GMR(H)$ curves (for the latter, we will consider for the angular dependence of the GMR only the case with the quadratic correction of Blaas et al.[51]).

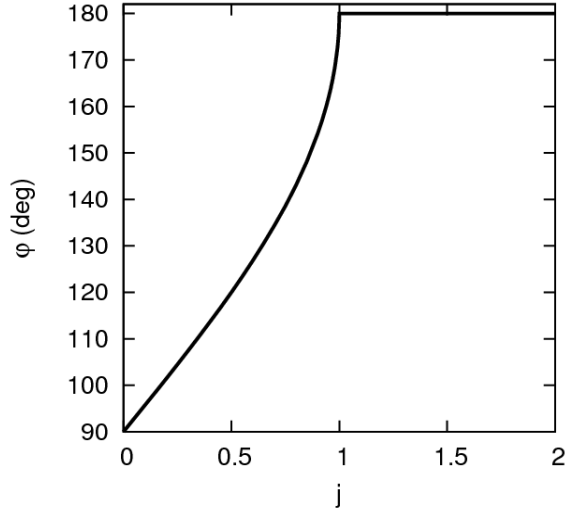

***Fig. 6*** *The equilibrium angle $\varphi$ between the magnetization vectors at H = 0 as a function of j = $J_1/J_2$ for the case of AF and orthogonal coupling simultaneously present.*

### *3.3.1. The case of* $j \geq 1$

If $j \geq 1$, there is no sudden jump in the magnetization curve: upon increasing the magnetic field, the initial (zero-field) antiparallel alignment goes over to the parallel alignment at $H = H_s$ continuously. It should also be noted that since the magnetization alignment is antiparallel for each value of j in the range $j \geq 1$, the remanence will be 0 for all values of j.

Figure 7 shows the magnetization curves (a) and the magnetoresistance curves (b) for $j \geq 1$. Increasing the value of *j* beyond 1 (Fig. 7a), the curves go over continuously into those which correspond to the pure antiferromagnetic coupling ($j = \infty$). If $j = 1$, the magnetization curve has an infinite slope at the (0,0) point. On the other hand, the slope of the *GMR(H)* curve is zero at (0,1) except when $j = 1$ because in this case the slope is infinite (Fig. 7b).

It should be noted here that for $j \geq 1$ (which is equivalent to $J_1 \geq J_2$, i.e., if the AF coupling is sufficiently strong with respect to the orthogonal coupling) and if there is no magnetic anisotropy, both the *M(H)* and *GMR(H)* curves remain free of hysteresis as was also for the case of pure AF coupling. This is because for $j \geq 1$ the magnetization alignement is antiparallel at $H = 0$ and starting to increase the magnetic field in either direction results in the same tilt of the two magnetization vectors towards the field direction..

It can be established, furthermore, from Fig. 7 that for the combined case of AF and orthogonal coupling, as $j \to \infty$ (more and more dominating AF coupling), both the *M(H)* and *GMR(H)* curves approach the corresponding curves derived for the pure AF coupling case (Fig. 3).

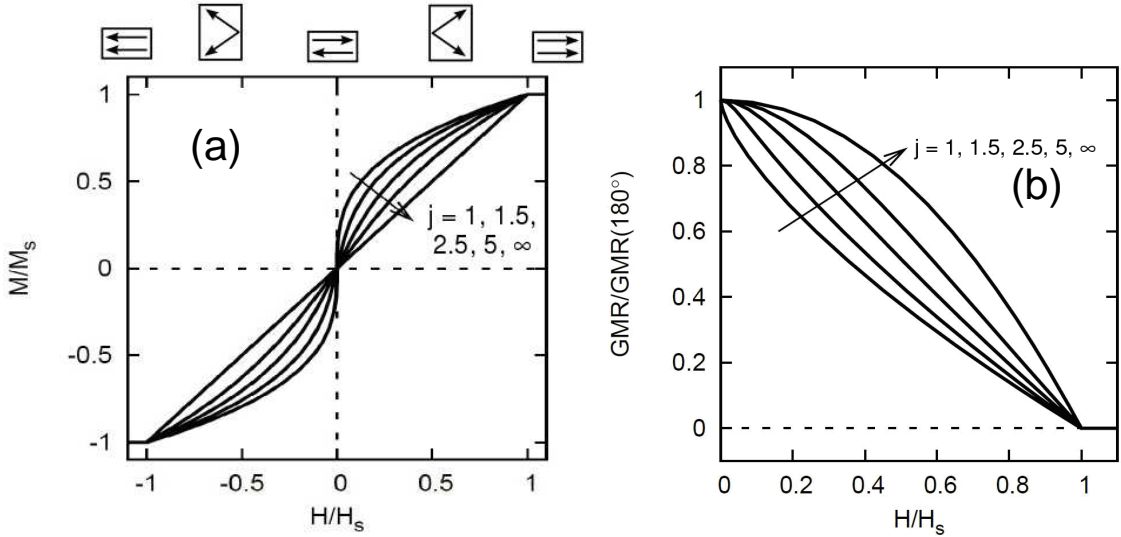

***Fig. 7*** *The magnetization curves (a) and the GMR(H) curves (b) for the case of AF and orthogonal coupling simultaneously present if $j \geq 1$. The arrows indicate the direction of the increase of the j values. The case $j = \infty$ means pure AF coupling. The arrows in the small boxes show the alignment of the magnetization vectors. In (b), the GMR(H) curves were calculated by taking into account also the quadratic correction[51] to the relation GMR($\varphi$) $\propto$ (1 - cos $\varphi$).*

### 3.3.2. The case of $j < 1$

If $j < 1$ (i.e., if the AF coupling is not sufficiently strong with respect to the orthogonal coupling), the magnetization process is qualitatively similar to the pure orthogonal case. The $M(H)$ and $GMR(H)$ curves are shown in Figs. 8a and 8b, respectively. As the value of $j$ increases from 0 towards 1, the relative remanence of the magnetization reduces continuously from $\frac{\sqrt{2}}{2}$ (~ 0.7071) to 0. The $GMR(H)$ curves exhibit the same shape (sharp peak at $H = 0$, downward curvature with increasing field) as was found for the pure orthogonal coupling (Fig. 5b). The magnitude of the GMR is the possible maximum as $j$ approaches 1 where the AF coupling dominates with perfect antiparallel alignment and it reduces to about 0.5 for $j = 0$ (pure orthogonal coupling, alignment with 90º). It is noted finally that hysteresis-free $M(H)$ and $GMR(H)$ curves are obtained also for $j < 1$.

It can be established from Fig. 8 that for the case of combined of AF and orthogonal coupling, as $j \to 0$ (more and more dominating orthogonal coupling), both the $M(H)$ and $GMR(H)$ curves recover the case of pure orthogonal coupling (Fig. 5).

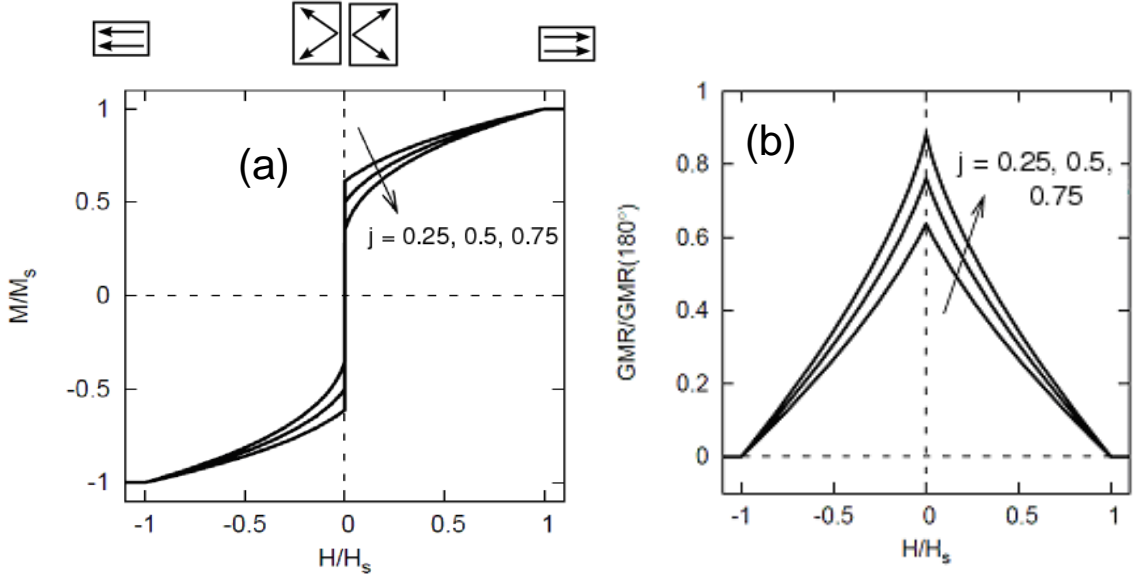

***Fig. 8*** *The magnetization curves (a) and the GMR(H) curves (b) for the case of AF and orthogonal coupling simultaneously present for some values of $j < 1$. In (a), the remanence $M_r$ gets smaller if we increase $j$ and at $j = 1$, the remanence vanishes. In (b), the height of the GMR peak increases if we increase $j$ and the GMR value at $j = 1$ is the possible maximum corresponding to that of pure AF coupling. The arrows in the small boxes show the alignment of the magnetization vectors. In (b), the GMR(H) curves were calculated by taking into account also the quadratic correction[51] to the relation $GMR(\varphi) \propto (1 - \cos \varphi)$.*

## 4. SUMMARY

In this work, we have modeled the field dependence of the magnetization and the *GMR* of FM/NM multilayers in the simplifying approach of a trilayer $FM_1$/NM/$FM_2$ with the two FM layers being identical. We assumed a single-domain state for each magnetic layer and that the magnetization vectors rotate coherently in the plane of the layers.

In agreement with most theoretical and experimental works, the angular dependence of the *GMR* was assumed to be described fairly well with the relation $GMR(\varphi) \propto (1 - \cos \varphi)$. By using this functional form, the *GMR*(*H*) curves were derived after establishing the field dependence of the equilibrium angles between the magnetizations of the two layers by minimizing the total magnetic energy. It was also found that a theoretically calculated small quadratic correction[51] to the relation $GMR(\varphi) \propto (1 - \cos \varphi)$ changed only slightly the calculated *GMR*(*H*) curves.

We succeeded in determining the *M*(*H*) and *GMR*(*H*) curves for the case of
(i)     pure AF coupling,
(ii)    orthogonal coupling and
(iii)   simultaneous presence of the AF and orthogonal coupling
for the full range of normalized parameter variables.

Some of our results could be compared with earlier works,[21-33] and these agree very well. For more complex combinations, the limiting cases yielded the results derived previously for simpler cases.

These calculations should help us better understand the magnetoresistance behavior of real FM/NM multilayers by a comparison of calculated and experimental *GMR*(*H*) curves.

It is noted that the results of modeling by considering also the presence of a unixial magnetic anisotropy in addition to the above treated couplings will be described in a subsequent paper.

**Acknowledgements** This work was supported by the Hungarian Scientific Research Fund through grant OTKA K 75008.